\documentclass[twocolumn,superscriptaddress,pra,showpacs]{revtex4}

\begin{document}
\title{Double dot chain as a macroscopic quantum bit}
\author{Ferdinando de Pasquale}
\email{ferdinando.depasquale@roma1.infn.it} \affiliation{INFM Center
for Statistical Mechanics and Complexity} \affiliation{Dipartimento
di Fisica, Universit\`{a} di Roma La Sapienza, Piazzale A. Moro 2,
00185 Roma, Italy}

\author{Gian Luca Giorgi}
\affiliation{INFM Center for Statistical Mechanics and Complexity}
\affiliation{Dipartimento di Fisica, Universit\`{a} di Roma La
Sapienza, Piazzale A. Moro 2, 00185 Roma, Italy}
\author{Simone Paganelli}
\affiliation{Dipartimento di Fisica, Universit\`{a} di Roma La
Sapienza, Piazzale A. Moro 2, 00185 Roma, Italy}
\affiliation{Dipartimento di Fisica, Universit\`{a} di Bologna, Via
Irnerio 46, I-40126, Bologna, Italy}


\begin{abstract}
We consider an array of $N$ quantum dot pairs interacting via
Coulomb interaction between adjacent dots and hopping inside each
pair. We show that at the first order in the ratio of hopping and
interaction amplitudes, the array maps in an effective two level
system with energy separation becoming exponentially small in the
macroscopic (large $N$) limit. Decoherence at zero temperature is
studied in the limit of weak coupling with phonons. In this case the
macroscopic limit is robust with respect to decoherence. Some
possible applications in quantum information processing are
discussed.
\end{abstract}
\maketitle

\section{Introduction}

In recent years a lot of attention has been devoted to the existence
of quantum superposition states in macroscopic systems. The first
suggestion to understand this phenomenon is due to Schr\"{o}dinger
\cite{schrodinger} who introduced the paradox of the cat in quantum
superposition between live and death, strongly stressing the
different behavior of the quantum world with respect to the human
experience. It is commonly accepted that quantum behavior vanishes
as the system size increases. Remarkable exceptions are quantum
systems which undergo a phase transition, as superconductors and
superfluids. A quantum superposition of mesoscopic states has been
observed in SQUID devices \cite{nakamura} and seems to be a
promising tool for the realization of a quantum computer.

Quantum macroscopic states are expected to be robust with respect to
decoherence and thus ideal candidates for quantum information
storage. Moreover, it is well known that a quantum system which
undergoes a phase transition lives in one of a particular set of
states, for a time which becomes infinitely large in the limit of
large system size. In particular, if the ground state is twofold
degenerate, one can associate these states to a macroscopic quantum
bit. The availability of macroscopic quantum bits is relevant for
quantum information processing, as shown, for instance, in Ref.
\cite{loss}, in the case of spin clusters. Here a new application in
teleportation processes is also shown.

Decoherence of a single qubit has been extensively studied
\cite{zurek,zurek 2,brune,walls,collett,paganel,giulini}. One of the
most relevant causes of decoherence is the coupling with a bosonic
bath \cite
{caldeira,anastopoulos,privman,palma,brandes,leggett,weiss} whose
effects are relevant also at zero temperature.

In the present paper we investigate the coherence of an array of $N$
double quantum dots coupled through Coulomb interaction in order to
show that such system is a suitable candidate as a macroscopic
qubit. The first step is to show that in the long time limit the
array behaves as a two level system with energy separation which
vanishes for large $N$. The model is exactly equivalent to a
one-dimensional antiferromagnetic Ising model in a transverse field.
The system is characterized by two equivalent charge configurations
and in the macroscopic limit the ordering at zero temperature
implies a separation of phase space in two regions around each of
the degenerate configurations. However, for a finite size system,
hopping induces oscillations between the two configurations. We
associate to this behavior a macroscopic quantum bit. The
antiferromagnetic Ising model in a transverse field has been studied
since the pioneering work of Bethe \cite {bethe,mattis,pfeuty}, and
has received recently a renewed attention as a model for quantum
computation \cite{cirac,fazio,osborne}. We found convenient to
introduce a simple approximation which makes transparent how the two
level behavior appears asymptotically.

The study of decoherence in such a system is analogous to
decoherence in a quantum register \cite{reina}. We show that, at
least in the weak coupling and zero temperature limit and for a
three-dimensional environment, the system exhibits a robustness
growing with the size of the array.

Decoherence with respect to phonons of a single two level system has
been studied with various methods (see reviews by Leggett \emph{{et
al.}} \cite{leggett} and Weiss \cite{weiss}). We found however the
resolvent method \cite{fujita,schotte}, already introduced in the
discussion of electron-phonon interaction problems \cite{suzuki},
convenient to obtain results at zero temperature in the double dot
chain.

The paper is organized as follows. In section \ref{2} we introduce
the model of the double dot chain and its interaction with a phonon
bath. The introduction of the resolvent method to discuss
decoherence will be the argument of section \ref{II}. In section
\ref{III} we shall apply the same method to show that the double dot
chain, in the limit $w/U<<1$, being $w$ the hopping amplitude
between dots inside each pair and $U$ the Coulomb interaction
between different pairs, behaves as an effective two level system
with energy separation decreasing exponentially with $N$. In section
\ref{IV} we study decoherence in our system in the approximation
introduced above. Finally, section \ref{V} is devoted to
conclusions. In appendix we review, inside the present
approximation, decoherence effects for a single dot pair.

\section{The model\label{2}}

In a previous work \cite{0312112} we proposed an array of few
coupled quantum dot pairs as a channel for teleportation. We want
here to show the extension of the model to a case of $N$ pairs and
discuss the robustness of the system with respect to decoherence due
to interaction with an external phonon bath at zero temperature. We
expect that the extensive character of the interaction will increase
decoherence while the macroscopic nature of the first two energy
states will enhance the robustness of the system. It will be shown
that the latter feature prevails. Neglecting spin effects, the
double dot array is characterized by the Hamiltonian
\begin{equation}
H_{S}=U\sum_{l=1}^{N-1}\sum_{\alpha =1}^{2}n_{l,\alpha
}n_{l+1,\alpha }-w\sum_{l=1}^{N}\left( c_{l,1}^{\dagger
}c_{l,2}+h.c.\right),
\end{equation}
where $c_{l,\alpha }^{\dag }$ creates an electron on the l (th) dot on the $%
\alpha $ (th) row of the array and $n_{l,\alpha}=c_{l,\alpha }^{\dag
}c_{l,\alpha }$.

To extend the teleportation scheme described in \cite{0312112}, we
introduce an initial superposition of two spin configurations of
zero potential energy of $N$ pairs:
\begin{equation}
\left| S\right\rangle =\alpha \left| \Phi \right\rangle +\beta
\left| \Psi \right\rangle,  \label{esse}
\end{equation}
where$\ \left| \Phi \right\rangle =\left| \downarrow ,\uparrow
,\downarrow ,\uparrow ,....\uparrow \right\rangle $ and $\left| \Psi
\right\rangle =\left| \uparrow ,\downarrow ,\uparrow ,\downarrow
,....\downarrow \right\rangle $. Let us consider an initial system
of $N-1$ double quantum dots with hopping inside any pair and
without Coulomb interaction. The ground state of this system is
represented by the tensor product of $\left( \left| \uparrow
\right\rangle +\left| \downarrow \right\rangle \right) /\sqrt{2}$
for each pair. By an adiabatic switching of electrostatic repulsion
between adjacent pairs, the system is driven in its new ground
state, which, for $w/U<<1$, is well approximated by $\left( \left|
\Phi _{N-1}\right\rangle +\left| \Psi
_{N-1}\right\rangle \right) /\sqrt{2}$. $\left| \Phi _{N-1}\right\rangle $ ($%
\left| \Psi _{N-1}\right\rangle $) is the same state than $\left|
\Phi \right\rangle $ ($\left| \Psi \right\rangle $), defined on
$N-1$ sites. The state $\left| S\right\rangle $ is obtained
considering an extra double dot (as usual called Alice) in a
superposition state $\alpha \left| \uparrow \right\rangle +\beta
\left| \downarrow \right\rangle $ and its interaction with the first
pair of $\left( \left| \Phi _{N-1}\right\rangle +\left| \Psi
_{N-1}\right\rangle \right) $. If the interaction is adiabatically
switched on again, then the system is driven in a state close to
$\left| S\right\rangle $. A proper manipulation of system parameters
permits to transfer the information, i.e. $\alpha $ and $\beta $,
encoded previously by Alice, to the last double dot (Bob)
\cite{0312112}.

The model described above is suitable to be represented by a spin
Hamiltonian through the mapping $\sigma _{i}^{z}=(n_{l,1}-n_{l,2})$
and $\sigma _{l}^{x}=(c_{l,1}^{\dag }c_{l,2}+h.c.)$.  This picture
is useful to study the decoherence effects induced by  the
interaction with a phonon bath. In the spin representation the
overall Hamiltonian becomes
\begin{eqnarray}
H &=&H_{S}+H_{B}+H_{SB}, \\
H_{S} &=&-w\sum_{l}\sigma _{l}^{x}+\frac{U}{2}\sum_{l}\left( \sigma
_{l}^{z}\sigma _{l+1}^{z}+1\right), \\
H_{B} &=&\sum_{\mathbf{q}}\omega _{\mathbf{q}}a_{\mathbf{q}}^{\dagger }a_{\mathbf{q}}, \\
H_{SB} &=&\sum_{\mathbf{q},l}g_{\mathbf{q}}n_{l}e^{iq \cos \theta
l}\left( a_{\mathbf{q}}^{\dagger }+a_{-\mathbf{q}}\right),
\label{accasb}
\end{eqnarray}
where a mapping between the spin states $\left| \uparrow
\right\rangle $ and $\left| \downarrow \right\rangle $ and the
charge states $\left|
1,0\right\rangle $ and $\left| 0,1\right\rangle $ has been performed and $%
n_{l}=\left( \sigma _{l}^{z}+1\right) /2$. We indicate with $\theta$
the angle between the phonon mode $\mathbf{q}$ and the  the dot
chain direction. This notation is useful for describing a generic
$d$-dimensional environment coupled with a one-dimensional system.
The constant $g_{\mathbf{q}}$ represents the coupling of the dot
charge with the mode $\mathbf{q}$. The explicit mathematical
expression for $g_{\mathbf{q}}$ depends on the specific
configuration of the system and the type of interaction. In Ref.
\cite{fedorov} the explicit form of $g_{q}$ in some remarkable case
is given.

\section{Resolvent method for weak coupling decoherence\label{II}}

Decoherence at zero temperature is studied using the resolvent
method. At the initial time $t=0$ system and bath are decoupled:
$\left| \Xi \left( t=0\right) \right\rangle =\left| S\right\rangle
\otimes \left| 0\right\rangle $ where $\left| 0\right\rangle $ is
the vacuum phonon state.

The time evolution of the state $\left| \Xi \left( t\right)
\right\rangle =\exp \left( -iHt\right) \left| \Xi \left( t=0\right)
\right\rangle $\ is studied in terms of the complex Laplace
transform defined as
\begin{equation}
\left| \Xi \left( \omega \right) \right\rangle = i \lim_{\delta
\rightarrow 0} \int_{0}^{\infty }e^{i\omega t-\delta t}\left| \Xi
\left( t\right) \right\rangle dt.
\end{equation}
The resolvent method allows to write
\begin{equation}
\left| \Xi \left( \omega \right) \right\rangle =\frac{1}{\omega
-H}\left| \Xi \left( t=0\right) \right\rangle.  \label{res}
\end{equation}
Using the identity
\begin{equation}
\frac{1}{\omega -H}=\frac{1}{\omega -H_{0}}+\frac{1}{\omega -H_{0}}H_{I}%
\frac{1}{\omega -H}  \label{omega}
\end{equation}
and performing a projection on the vacuum phonon state, we define a
new system state $\left| \Phi _{S}\left( \omega \right)
\right\rangle =\left\langle 0|\Xi \left( \omega \right)
\right\rangle $ that obeys to the evolution equation
\begin{eqnarray}
\left| \Phi _{S}\left( \omega \right) \right\rangle =\frac{1}{\omega
-H_{S}}
\left| \Phi _{S}\left( t=0\right) \right\rangle +\nonumber\\
\left\langle 0\right| \frac{1}{\omega -H_{S}}H_{SB}\frac{1}{\omega
-H}\left| \Xi \left( t=0\right) \right\rangle.  \label{fis}
\end{eqnarray}
Here the bath ground state energy is set to zero and
$H_{0}=H_{S}+H_{B}$ and $H_{I}=H_{SB}$.

In the weak coupling \ limit only corrections to the imaginary part of $%
\left| \Phi _{S}\left( \omega \right) \right\rangle $ will be taken
into account. We first perform an iteration inside Eq. \ref{fis}
replacing $\left( \omega -H\right) ^{-1}$ with the right hand side
of Eq. \ref{omega}, and then introduce a complete set of
intermediate phonon states:
\begin{widetext}
\begin{eqnarray}
\left| \Phi _{S}\left( \omega \right) \right\rangle &=&\frac{1}{\omega -H_{S}%
}\left| \Phi _{S}\left( t=0\right) \right\rangle + \left\langle
0\right| \frac{1}{\omega -H_{S}}H_{SB}\frac{1}{\omega
-H_{S}-H_{B}}\left| \Xi \left(
t=0\right) \right\rangle   \nonumber \\
&&+\sum_{k}\left\langle 0\right| \frac{1}{\omega
-H_{S}}H_{SB}\frac{1}{\omega
-H_{S}-H_{B}}H_{SB}\left| k\right\rangle \left\langle k\right| \frac{1}{%
\omega -H}\left| \Xi \left( t=0\right) \right\rangle.  \label{self}
\end{eqnarray}
\end{widetext}
In a perturbative approach, terms involving powers of $g_{q}$ are
small and can be neglected, unless self-energy contributes appear.
In the latter case, a not negligible imaginary part can arise
performing the sum over $q$ in the continuous limit.

The contribution involving self-energy in the sum corresponds to
$k=0$, since $\left\langle 0\right| \left( \omega -H\right)
^{-1}\left| \Xi \left( t=0\right) \right\rangle $ is exactly $\left|
\Phi _{S}\left( \omega \right)
\right\rangle $. Then, all other linear and quadratic contributions in $%
H_{SB}$ will be neglected. Hence, Eq. \ref{self} becomes
\begin{equation}\label{eqn:tnoto}
\left(1-\frac{1}{\omega-H_S}G(H_S)\right)  \left| \Phi _{S}\left(
\omega \right) \right\rangle = \frac{1}{\omega -H_{S}} \left| \Phi
_{S}\left( t=0\right) \right\rangle,
\end{equation}
where
\begin{equation}
G\left( H_{S}\right) =\left\langle 0\right| H_{SB}\frac{1}{\omega
-H_{S}-H_{B}}H_{SB}\left| 0\right\rangle
\end{equation}
is the self-energy operator acting on the system subspace. The right
term of Eq.\ref{eqn:tnoto} describes the evolution of the
macroscopic state isolated from phonons. As we will show in section
\ref{III}, in the limit of $w/U\ll 1$, the macroscopic dot chain
behaves as a two level system oscillating between the $H_S$
asymptotic eigenstates $\left| \pm \right\rangle=2^{-1/2}
(\left|\Phi \right\rangle \pm \left|\Psi \right\rangle)$
with energies $E_{\pm }$. So, Eq.\ref{eqn:tnoto} becomes
\begin{widetext}
\begin{equation}\label{eqn:tnoto1}
\left(1-\frac{1}{\omega-H_S}G(H_S)\right)  \left| \Phi _{S}\left(
\omega \right) \right\rangle = \frac{1}{\omega -E_{+}}\left| +
\right\rangle \left\langle +|\Phi _{S}\left( t=0\right)
\right\rangle+\frac{1}{\omega -E_{-}} \left| -
\right\rangle\left\langle -|\Phi _{S}\left( t=0\right)
\right\rangle.
\end{equation}
\end{widetext}
Noting that the operator $G(H_S)$ maps the subspace spanned by
$\left| \pm \right\rangle$ into itself, it is possible to reduce
Eq.\ref{eqn:tnoto1} in terms of two coupled equations:
\begin{eqnarray}
\left( \omega -E_{+}-G^{++}\right) \left\langle +|\Phi _{S}\left(
\omega \right) \right\rangle -G^{+-}\left\langle -|\Phi _{S}\left(
\omega \right) \right\rangle &=&\nonumber\\\left\langle +|\Phi
_{S}\left( t=0\right) \right\rangle,
\end{eqnarray}
\begin{eqnarray}
\left( \omega -E_{-}-G^{--}\right) \left\langle -|\Phi _{S}\left(
\omega \right) \right\rangle -G^{-+}\left\langle +|\Phi _{S}\left(
\omega \right) \right\rangle &=&\nonumber\\\left\langle -|\Phi
_{S}\left( t=0\right) \right\rangle,
\end{eqnarray}
where $G^{\pm \pm }=\left\langle \pm \right| G\left| \pm
\right\rangle $.

To the leading order in the system-bath coupling, we obtain
\begin{eqnarray}
\left\langle +|\Phi _{S}\left( \omega \right) \right\rangle &=&\frac{1}{%
\omega -E_{+}-G^{++}}\left\langle +|\Phi _{S}\left( t=0\right)
\right\rangle,
\\
\left\langle -|\Phi _{S}\left( \omega \right) \right\rangle &=&\frac{1}{%
\omega -E_{-}-G^{--}}\left\langle -|\Phi _{S}\left( t=0\right)
\right\rangle.
\end{eqnarray}

The solution in the time domain is obtained assuming first the
correction introduced by the matrix elements of $G$ as negligible,
and then calculating the latter in in $\omega=E_+$ or $\omega=E_-$.

For instance, the integral
\begin{eqnarray}
\int_{C}\frac{e^{-i\omega t}}{\omega -E_{+}-G^{++}}d\omega \nonumber
\end{eqnarray}
is calculted assuming first $G^{++}=0$, obtaining for the pole
$\omega =E_{+} $, and then substituting this value inside $G^{++}$,
which depends on $\omega $. After, the principal value of $G^{++}$
will be ignored, and only the imaginary part will matter. We
compared, in appendix, the results of our approximation, with those
known for single quantum dot pairs.

\section{Double dot array evolution\label{III}}

As first step we calculate the evolution of the system when it is
decoupled from the bath. We find convenient to explicitly solve the
evolution from an initial state corresponding respectively to
$\left| \Phi \right\rangle $ or $\left| \Psi \right\rangle $
introduced in Eq. \ref{esse}.

We distinguish in the system Hamiltonian the hopping term $%
H_{I}=-w\sum_{l}\sigma _{l}^{x}$ from the potential energy
$H_{0}=U/2\left[ \sum_{l}\left( \sigma _{l}^{z}\sigma
_{l+1}^{z}+1\right) \right] $. This is the antiferromagnetic version
of the well known one-dimensional Ising Model in a transverse field
\cite{sachdev}. It is worth noting that the absence of periodic
boundary conditions implies a relaxation mechanism of an initially
ordered state where a single domain wall propagates between the two
end points of the array. This feature makes a difference in the
excitation spectrum which is relevant for an array of finite size.

Applying $H_{I}$\ on $\left| \Phi \left( t=0\right) \right\rangle $
the system is driven in a new configuration labeled as $\left| \Phi
_{1}\left( t=0\right) \right\rangle $. The action of $H_{I}$
generates a sum of states each of them differentiates from $\left|
\Phi \left( t=0\right) \right\rangle $ due to one spin flip in a
different place along the array. Here it is important to note that
flips on the first and the last qubit put the system in a state with
Coulomb energy $U$, while all intermediate transitions lead to a
state with a $2U$ electrostatic energy. In the limit of $U$ large
with respect to $w$, we shall neglect all configurations involving
intermediate states with energy greater than $U$.

In each step of a repeated application of $H_{I}$ it is possible to
go towards new configurations or to come back. Then, for $n>0$, we
write
\begin{equation}
H_{I}\left| \Phi _{n}\left( t=0\right) \right\rangle =-w\left[
\left| \Phi _{n-1}\left( t=0\right) \right\rangle +\left| \Phi
_{n+1}\left( t=0\right) \right\rangle \right].
\end{equation}
After $N$ steps the system reaches $\left| \Psi \right\rangle $ and after $%
2N $ steps it comes back to the initial configuration. Defining
$\left| \Phi _{N}\right\rangle =\left| \Psi \right\rangle $ and
$\left| \Phi _{0}\right\rangle =\left| \Phi _{2N}\right\rangle
=\left| \Phi \right\rangle $, and taking into account the time
evolution we obtain
\begin{widetext}
\begin{equation}
\left( \omega -U\right) \left| \Phi _{n}\left( \omega \right)
\right\rangle =\left| \Phi _{n}\left( t=0\right) \right\rangle
-w\left[ \left| \Phi _{n-1}\left( \omega \right) \right\rangle
+\left| \Phi _{n+1}\left( \omega \right) \right\rangle \right]
-U\left( \delta _{n,0}+\delta _{n,N}\right) \left| \Phi _{n}\left(
\omega \right) \right\rangle.  \label{fienne}
\end{equation}
\end{widetext} The system is solved by means of the discrete Fourier
transform defined as
\begin{eqnarray}
\left| \widetilde{\Phi }_{k}\left( \omega \right) \right\rangle &=&\frac{1}{%
\sqrt{2N}}\sum_{n=0}^{2N-1}\left| \Phi _{n}\left( \omega \right)
\right\rangle e^{ink},  \nonumber \\
\left| \Phi _{n}\left( \omega \right) \right\rangle &=&\frac{1}{\sqrt{2N}}%
\sum_{k=0}^{2N-1}\left| \widetilde{\Phi }_{k}\left( \omega \right)
\right\rangle e^{-ink}.
\end{eqnarray}
As a consequence of periodicity conditions, $k=\frac{2\pi }{2N}n$ being $%
n=0,1,2,...,2N-1$.\

From \ Eq. \ref{fienne} follows
\begin{eqnarray}
\left[ \omega -U+2w\cos k\right] \left| \widetilde{\Phi }_{k}\left(
\omega \right) \right\rangle =\left| \widetilde{\Phi }_{k}\left(
t=0\right) \right\rangle \nonumber\\-\frac{U}{\sqrt{2N}}\left(
\left| \Phi _{0}\left( \omega \right) \right\rangle +e^{iNk}\left|
\Phi _{N}\left( \omega \right) \right\rangle \right).
\end{eqnarray}

It's now possible to extract two equations connecting $\left| \Phi
_{0}\right\rangle $ to $\left| \Phi _{N}\right\rangle $:
\begin{equation}
\left| \Phi _{0}\left( \omega \right) \right\rangle =\frac{\left[
1+B_{0}\left( \omega \right) \right] \left| A_{0}\left( \omega
\right) \right\rangle -B_{N}\left( \omega \right) \left| A_{N}\left(
\omega \right) \right\rangle }{\left[ 1+B_{0}\left( \omega \right)
\right] ^{2}-B_{N}^{2}\left( \omega \right) } ,
\end{equation}
\begin{equation}
\left| \Phi _{N}\left( \omega \right) \right\rangle =\frac{\left[
1+B_{0}\left( \omega \right) \right] \left| A_{N}\left( \omega
\right) \right\rangle -B_{N}\left( \omega \right) \left| A_{0}\left(
\omega \right) \right\rangle }{\left( 1+B_{0}\left( \omega \right)
\right) ^{2}-B_{N}^{2}\left( \omega \right) },
\end{equation}

where
\begin{equation}
\left| A_{n}\left( \omega \right) \right\rangle =\frac{1}{\sqrt{2N}}%
\sum_{k=0}^{2\pi \left( \frac{2N-1}{2N}\right) }\frac{e^{-ink}\left|
\widetilde{\Phi }_{k}\left( t=0\right) \right\rangle }{\omega
-U+2w\cos k},
\end{equation}

\begin{equation}
B_{n}\left( \omega \right) =\frac{1}{2N}\frac{U}{\omega -U}\sum_{q=0}^{2N-1}%
\frac{e^{-in\frac{\pi }{N}q}}{1-a\left( \omega \right) \cos
\frac{\pi }{N}q},
\end{equation}
with $a\left( \omega \right) =2w/\left( U-\omega \right) $ and noting that $%
B_{N}=B_{-N}$.

The asymptotic behavior is determined by values of $\omega $ close to zero. Then $%
a\left( \omega \right) <<1$ and the denominator of $B_{n}\left(
\omega \right) $ reads as geometric series:
\begin{equation}
B_{n}\left( \omega \right) =\frac{1}{2N}\frac{U}{\omega -U}%
\sum_{q=0}^{2N-1}e^{-in\frac{\pi }{N}q}\sum_{l=0}^{\infty
}a^{l}\left( \omega \right) \cos ^{l}\frac{\pi }{N}q,
\end{equation}
or
\begin{eqnarray}
B_{n}\left( \omega \right) =\frac{1}{2N}\frac{U}{\omega -U}%
\sum_{q=0}^{2N-1}\sum_{l=0}^{\infty }\sum_{m=0}^{l}%
{l \choose m}%
\left( \frac{a\left( \omega \right) }{2}\right) ^{l}\nonumber\\\times\exp \left[ i\frac{\pi }{%
N}\left( l-2m-n\right) q\right].
\end{eqnarray}
The sum over $q$ gives
\begin{equation}
B_{n}\left( \omega \right) = \frac{1}{2N}\frac{U}{\omega -U}%
\sum_{l=0}^{\infty }\sum_{m=0}^{l}%
{l \choose m}%
 \frac{a^l\left( \omega \right) }{2^l} \frac{1-e^{2i\pi
\left( l-2m-n\right) }}{1-e^{i\frac{\pi }{N}\left( l-2m-n \right)
}}.
\end{equation}
The condition for a nonvanishing $B_{n}\left( \omega \right) $ is
$\left(
l-2m-n\right) =2NK$, where $K$ is any integer between $-\infty $ and $%
+\infty $:
\begin{equation}
B_{n}\left( \omega \right) =\frac{U}{\omega -U}\sum_{l=0}^{\infty
}\sum_{m=0}^{l}\frac{l!}{m!\left( l-m\right) !} \frac{a^l\left(
\omega \right) }{2^l} \delta _{\left( l-2m-n\right) ,2NK}.
\end{equation}
or, using the Kronecker Delta function
\begin{equation}
B_{n}\left( \omega \right) =\frac{U}{\omega -U}\sum_{l=0}^{\infty
}\sum_{K=-\infty }^{\infty }\frac{l!}{\left(
\frac{l+n+2NK}{2}\right) !\left( \frac{l-n-2NK}{2}\right) !}
\frac{a^l\left( \omega \right) }{2^l} .  \label{exact}
\end{equation}
Since the coefficients of a Newton's binomial formula have to be
real and positive, in the limit $a\left( \omega \right) <<1$ we
obtain
\begin{equation}
B_{0}\left( \omega \right) \simeq \frac{U}{\omega -U}\left(
1+M\right),
\end{equation}
where
\begin{equation}
M=1-\frac{1}{2N}\sum_{q=0}^{2N-1}\frac{1}{1-\frac{2w}{U}\cos q}
\end{equation}
contains powers of $w/U$ and has to be calculated at the desired order in $q$%
, and
\begin{equation}
B_{N}\left( \omega \right) \simeq -\frac{1}{2^{N}}\left(
\frac{2w}{U}\right) ^{N}.
\end{equation}
Here we note that the last contribution cannot be ignored because it
gives
rise to the energy separation between $\left| \Phi _{0}\right\rangle $ and $%
\left| \Phi _{N}\right\rangle $.

Furthermore we obtain
\begin{equation}
\left| A_{0}\left( \omega \right) \right\rangle \simeq
\frac{1}{U}\left| \Phi _{0}\left( t=0\right) \right\rangle
\end{equation}
and
\begin{equation}
\left| A_{N}\left( \omega \right) \right\rangle \simeq
\frac{1}{U}\left| \Phi _{N}\left( t=0\right) \right\rangle.
\end{equation}
As a result, after an inverse Laplace transform, we get, apart from
corrections containing powers of $w/U$,
\begin{equation}
\left| \Phi _{0}\left( t\right) \right\rangle =e^{iMUt}\left[ \left|
\Phi _{0}\left( 0\right) \right\rangle \cos \Delta t
 -i\left| \Phi _{N}\left( 0\right) \right\rangle \sin
\Delta t\right]
\end{equation}
and
\begin{equation}
\left| \Phi _{N}\left( t\right) \right\rangle =e^{iMUt}\left[ \left|
\Phi _{N}\left( 0\right) \right\rangle \cos \Delta t-i\left| \Phi
_{0}\left( 0\right) \right\rangle \sin \Delta t\right],
\end{equation}
having introduced the energy gap
\begin{equation}
\Delta =2w\left( 2w/U\right) ^{N-1}.  \label{delta}
\end{equation}

We eventually obtain the long time behavior of a two level system
with energy separation exponentially vanishing in the large $N$
limit. Actually, in Ref. \cite{mattis} (see equation (3.32c)) the
eigenvalue of Eq. \ref {delta} was derived. On the basis of this
result the phenomenon of asymptotic degeneracy was established and
shown to be directly related to the appearance of the ordered phase
in the large $N$ limit.

\section{Double dot array decoherence in the long time limit\label{IV}}

According to the previous analysis we can limit ourself to consider
only the first two states $\left| \pm \right\rangle $ of the array.
The decoherence rate will be however modified by the extensive
interaction with the bath.

We have to calculate the matrix elements of $G\left( H_{S}\right) $
in the subspace of $\left| +\right\rangle $ and $\left|
-\right\rangle $ taking into account the particular system-bath
interaction $H_{SB}$ defined in Eq. \ref{accasb}. Here
\begin{equation}
G\left( H_{S}\right) =\sum_{\mathbf{q},l,l^{\prime }}e^{i q \cos
\theta \left( l-l^{\prime }\right)
}\left| g_{\mathbf{q}}\right| ^{2}n_{l^{\prime }}\frac{1}{\omega -H_{S}-\omega _{\mathbf{q}}}%
n_{l},
\end{equation}
where the sum over $l,l^{\prime }$ runs over the array sites where
electrons are present.

We choose the basis elements $\left| +\right\rangle $ and $\left|
-\right\rangle $ defined respectively as the sum and the difference of $%
\left| \Phi \right\rangle $ and $\left| \Psi \right\rangle $.

We introduce the form factor $\Lambda _{q\cos\theta}$ defined
through
\begin{equation}
\sum_{l}n_{l}e^{i q \cos\theta l}\left| \Phi \right\rangle =\Lambda
_{q \cos\theta}\left| \Phi \right\rangle
\end{equation}
or
\begin{equation}
\sum_{l}n_{l} e^{i q \cos\theta l} \left| \Psi \right\rangle
=e^{iq}\Lambda _{q \cos\theta}\left| \Psi \right\rangle
\end{equation}.
Explicitly, $\Lambda _{q \cos\theta}= \left( 1-e^{i2 q \cos\theta
N}\right) /\left( 1-e^{i2q \cos\theta}\right) $.

The matrix elements of the self-energy operator are
\begin{widetext}
\begin{eqnarray}
G^{++} &=&\sum_{\mathbf{q}}\left| g_{\mathbf{q}}\right| ^{2}\left|
\Lambda _{q \cos\theta}\right| ^{2}
\left[ \frac{\cos ^{2}\frac{q \cos\theta}{2}}{\omega -E_{+}-\omega _{\mathbf{q}}}+\frac{\sin ^{2}%
\frac{q \cos\theta}{2}}{\omega -E_{-}-\omega _{\mathbf{q}}}\right],  \\
G^{--} &=&\sum_{\mathbf{q}}\left| g_{\mathbf{q}}\right| ^{2}\left|
\Lambda _{q \cos\theta}\right| ^{2}
\left[ \frac{\cos ^{2}\frac{q \cos\theta}{2}}{\omega -E_{-}-\omega _{\mathbf{q}}}+\frac{\sin ^{2}%
\frac{q \cos\theta}{2}}{\omega -E_{+}-\omega _{\mathbf{q}}}\right],\\
G^{-+} &=&(G^{+-})^*=i\sum_{\mathbf{q}}\left| g_{\mathbf{q}}\right|
^{2}\left| \Lambda _{q \cos\theta}\right| ^{2}\cos \frac{q
\cos\theta}{2}\sin \frac{q \cos\theta}{2}\left[ \frac{1}{\omega -E_{+}-\omega _{\mathbf{q}}%
}-\frac{1}{\omega -E_{-}-\omega _{\mathbf{q}}}\right].
\end{eqnarray}
We find convenient the introduction of the following generalized
densities of states
\begin{eqnarray}
\rho ^{+-}\left( \epsilon \right) &=&(\rho ^{-+})^*\left( \epsilon
\right)=-i\sum_{\mathbf{q}}\left| g_{\mathbf{q}}\right| ^{2}\left|
\Lambda _{q \cos\theta}\right| ^{2}\cos \frac{q \cos\theta}{2}\sin
\frac{q \cos\theta}{2}\delta
\left( \epsilon -\omega _{\mathbf{q}}\right), \\
\rho _{1}\left( \epsilon \right) &=&\sum_{\mathbf{q}}\left|
g_{\mathbf{q}}\right| ^{2}\left| \Lambda _{q \cos\theta}\right|
^{2}\cos ^{2}\frac{q \cos\theta}{2}\delta \left( \epsilon -\omega
_{\mathbf{q}}\right), \\
\rho _{2}\left( \epsilon \right) &=&\sum_{\mathbf{q}}\left|
g_{\mathbf{q}}\right| ^{2}\left| \Lambda _{q \cos\theta}\right|
^{2}\sin ^{2}\frac{q \cos\theta}{2}\delta \left( \epsilon -\omega
_{\mathbf{q}}\right),
\end{eqnarray}
from which follows
\begin{eqnarray}
G^{++} &=&\int d\epsilon \left[ \frac{\rho _{1}\left( \epsilon \right) }{%
\omega -E_{+}-\epsilon }+\frac{\rho _{2}\left( \epsilon \right)
}{\omega
-E_{-}-\epsilon }\right], \\G^{+-} &=&-i\int d\epsilon \rho ^{+-}\left( \epsilon \right) \left[ \frac{1}{%
\omega -E_{+}-\epsilon }-\frac{1}{\omega -E_{-}-\epsilon }\right], \\
G^{--} &=&\int d\epsilon \left[ \frac{\rho _{1}\left( \epsilon \right) }{%
\omega -E_{-}-\epsilon }+\frac{\rho _{2}\left( \epsilon \right)
}{\omega -E_{+}-\epsilon }\right].
\end{eqnarray}
\end{widetext}
The real part of $G$ gives a negligible contribution to the pole
location if compared with $E_{-}$ and $E_{+}$. Thus, assuming a
density of state
different from zero only for positive $\epsilon $, as in Ref. \cite{brandes}%
, the only non vanishing contribution is $\gamma =-%
\mathop{\rm Im}%
G^{--}$:
\begin{equation}
\gamma =-\pi \rho _{2}\left( \Delta \right),
\end{equation}
where $\Delta =E_{-}-E_{+}$ is the energy gap of the two level
system and is positive (being $\left| +\right\rangle $ the ground
state).

Then the solution for $\left\langle +|\Phi _{S}\left( t\right)
\right\rangle $ and $\left\langle -|\Phi _{S}\left( t\right)
\right\rangle $ is
\begin{eqnarray}
\left\langle +|\Phi _{S}\left( t\right) \right\rangle
&=&e^{iE_{+}t}\left\langle +|\Phi _{S}\left( t=0\right) \right\rangle, \\
\left\langle -|\Phi _{S}\left( \omega \right) \right\rangle
&=&e^{iE_{-}t-\gamma t}\left\langle -|\Phi _{S}\left( t=0\right)
\right\rangle.
\end{eqnarray}

As expected, the ground state is not affected by decoherence, while
the excited state relaxes. Damping is proportional to the density of
states calculated at the energy gap. The density of states is
however quite different from that of a single dot pair. Two
competitive effects appear. The first one is represented by the
presence of the form factor $\Lambda _{q \cos\theta}$ inside $\rho
_{2}$, which, in the large $N$ limit, increases the dephasing rate
by a factor proportional to $N^{2}$. The second, predominant, effect
to be considered is the expontential reduction with $N$ of the
energy separation.

For instance, in the simple case of $\left| g_{\mathbf{q}}\right| ^{2}=1/N$ and $%
\omega _{\mathbf{q}}=cq$ (longitudinal phonons)
\begin{equation}
\gamma \left( \Delta \right) \propto \int d \cos \theta d^{d}q\frac{\sin ^{2}qN}{\sin ^{2}q \cos\theta%
}\sin ^{2}\frac{q \cos\theta}{2}\delta \left( \Delta
-c^{2}q^{2}\right),
\end{equation}
where $d$ is the dimension of the bath and $c$ is the speed of
sound. If we compare this quantity with the system oscillation
frequency we obtain
\begin{equation}\label{}
\frac{\gamma \left( \Delta \right)}{\Delta}\propto N^{2}\Delta
^{d/2-1}.
\end{equation}
This result indicates that, for a phonon bath in three dimensions,
the macroscopic limit involves a growth of the robustness with
respect to decoherence.

\section{Conclusions\label{V}}

The existence of a macroscopic degenerate ground state is a general
feature of a system exhibiting a phase transition. It would be of
interest to exploit such degenerate states as elements of a
macroscopic qubit. Here we considered an array of $N$ interacting
dot pairs. We showed that, in the long time limit and neglecting
correction of order $w/U$, the system oscillates between the two
configurations characterized by zero electrostatic energy. Such a
system has a robustness which increases with the size as shown
calculating the decoherence due to a phonon bath at zero
temperature. Decoherence calculation has been performed in the
framework of an application of the resolvent method. It is however
important to note that a general conclusion about robustness of
macroscopic qubits should consider nonzero temperature effects and
other different mechanisms for daphasing in semiconductors (such as
cotunneling and background charge fluctuations).

As said, coherent manipulation of $\left| S\right\rangle $ is useful
in order to realize a support for quantum information transfer
allowing a solid state teleportation. The basic information
processing steps, i.e. initialization of the system, local gates and
readout are respectively
obtained by adiabatic variation of system parameters, oscillations between $%
\left| \Phi \right\rangle $ and $\left| \Psi \right\rangle $) and
local charge measurement. Moreover, it is worth to note that,
because of the analogy with spin clusters\ behavior, most of
considerations on quantum computing developed in Ref. \cite{loss}
should be valid also for our system, with the advantage that instead
of measuring spin states, we need to detect local charges. Due to
the asymptotic behavior of the effective hopping amplitude, such a
qubit requires gate times exponentially increasing with the size of
the system and the optimal chain length will be determined by the
specific application.

\renewcommand{\theequation}{A-\arabic{equation}}
\setcounter{equation}{0} 

\section*{APPENDIX A: DECOHERENCE RATE IN A DOUBLE QUANTUM DOT\label{App}}

We introduce a double quantum dot in contact with a bosonic bath
with Hamiltonian

\begin{eqnarray}
H &=&H_{S}+H_{B}+H_{SB},  \nonumber \\
H_{S} &=&\frac{\varepsilon }{2}\sigma _{z}+T\sigma _{x},  \nonumber \\
H_{B} &=&\sum_{q}\omega _{q}a_{q}^{\dagger }a_{q},  \nonumber \\
H_{SB} &=&\frac{1}{2}\sigma _{z}\sum_{q}g_{q}\left( a_{q}^{\dagger
}+a_{q}\right),
\end{eqnarray}
discussed in Ref. \cite{brandes}. Here a one-dimensional bath is
considered for simplicity. Labeling with $\left| L\right\rangle $
and $\left| R\right\rangle $ the eigenstates of $\sigma _{z}$ with
respective eigenvalues $+1$ and$-1$, the eigenstates of $H_{S}$ are
\begin{equation}
\left| \pm \right\rangle =\frac{1}{N_{\pm }}\left[ \pm 2T\left|
L\right\rangle +\left( \Delta \mp \varepsilon \right) \left| R\right\rangle %
\right],
\end{equation}
where $\Delta =\sqrt{\varepsilon ^{2}+4T^{2}}$ and $N_{\pm }=\sqrt{(\Delta {%
\mp }\varepsilon )^{2}+4T^{2}}$ while the respective eigenvalues are $%
\varepsilon _{{\pm }}=\pm \frac{1}{2}\Delta $.

By inversion we obtain
\begin{eqnarray}
\left| L\right\rangle &=&N_{+}\text{ }\frac{\Delta +\varepsilon }{4T\Delta }%
\left| +\right\rangle -N_{-}\text{ }\frac{\Delta -\varepsilon }{4T\Delta }%
\left| -\right\rangle, \\
\left| R\right\rangle &=&\text{ }\frac{N_{+}}{2\Delta }\left|
+\right\rangle +\text{ }\frac{N_{-}}{2\Delta }\left| -\right\rangle.
\end{eqnarray}

Eq. \ref{eqn:tnoto} has now to be solved using

\begin{equation}
G\left( H_{S}\right) =\frac{1}{4}\sum_{q}\left| g_{q}\right| ^{2}\sigma _{z}%
\frac{1}{\omega -\omega _{q}-H_{S}}\sigma _{z}.
\end{equation}
We need to calculate $\left\langle +\right| G\left( H_{S}\right)
\left| +\right\rangle $ and $\left\langle -\right| G\left(
H_{S}\right) \left| -\right\rangle $. Actually, obtaining $G^{++}$
will be enough, due the intrinsic robustness of the ground state
$\left| -\right\rangle $ \cite {preskill2} which implies that
$G^{--}$ has to be zero (this feature is easily checked in the
present formalism). To do it first we write $\left| +\right\rangle $
in the $\left| L,R\right\rangle $ basis, then apply $\sigma _{z}$,
come back in the $\left| \pm \right\rangle $ basis in order to
apply $(\omega -\omega _{q}-H_{S})^{-1}$, rewrite the new state through $%
\left| L,R\right\rangle $ to apply the second $\sigma _{z}$
operator, and finally re-express the result in terms of $\left| +\right\rangle $ and $%
\left| -\right\rangle $. The result is
\begin{widetext}
\begin{eqnarray}
G^{++}&=&\frac{1}{4}\sum_{q}\left| g_{q}\right| ^{2}\left[
\frac{1}{\omega -\omega _{q}-\frac{\Delta }{2}}\left(
\frac{\varepsilon }{\Delta }\right) ^{2}+\frac{1}{\omega -\omega
_{q}+\frac{\Delta }{2}}\left( \frac{\Delta -\varepsilon }{\Delta
}\right) ^{2}\right].
\end{eqnarray}
\end{widetext}
The sum over $q$ is performed as an integral through the
introduction of the
density of states $\rho $ which is assumed to be different from zero
only for positive values of its argument \cite{brandes}. The second
term inside the
square bracket gives the contribution to the imaginary part, which is $%
\gamma =-\left[ \pi T^{2}\rho \left( \Delta \right) \right] /\Delta
^{2}$. The evolution is thus
\begin{equation}\label{A7}
\left\langle +|\Phi _{S}\left( t\right) \right\rangle =\left\langle
+|\Phi
_{S}\left( t=0\right) \right\rangle e^{-i\frac{\Delta }{2}t}e^{-\pi \frac{%
T^{2}}{\Delta ^{2}}\rho \left( \Delta \right) t},
\end{equation}
\begin{equation}
\left\langle -|\Phi _{S}\left( t\right) \right\rangle =\left\langle
-|\Phi _{S}\left( t=0\right) \right\rangle e^{i\frac{\Delta }{2}t}.
\end{equation}
The density matrix in the basis $\left| \pm \right\rangle $ is then

\begin{equation}
\rho (t)=\left(
\begin{array}{cc}
\rho ^{++}\left( 0\right) e^{-2\gamma t} & \rho ^{-+}\left( 0\right)
e^{-\gamma t}e^{i\Delta t} \\
\rho ^{-+}\left( 0\right) e^{-\gamma t}e^{i\Delta t} & 1-\rho
^{++}\left( 0\right) e^{-2\gamma t}
\end{array}
\right).
\end{equation}
with the same dephasing rate obtained in \cite{brandes}, in the
regime of zero temperature, using markovian assumptions.

\end{document}